\documentclass[showpacs,aps,twocolumn]{revtex4-1}

\usepackage{bm}
\usepackage{amsmath}
\usepackage{graphicx}
\usepackage{subfigure}
\usepackage[usenames,dvipsnames]{color}
\definecolor{darkblue}{RGB}{0,0,196}
\usepackage[colorlinks=true,linkcolor=darkblue,citecolor=darkblue,urlcolor=darkblue]{hyperref}
\usepackage{setspace}
\usepackage{footmisc}
\usepackage[makeroom]{cancel}
\usepackage{url}
\usepackage{multirow}

\usepackage{color,soul}

\def\be{\begin{equation}}
\def\ee{\end{equation}}
\def\ba{\begin{eqnarray}}
\def\ea{\end{eqnarray}}

\usepackage{graphicx}
\usepackage{amsmath,bbm}
\usepackage{amssymb,bm}

\begin{document}

\title{Transport of $D^0$ meson in hadronic matter in the domain of Non-Extensive statistics}
\author{Aditya~Kumar~Singh}
\author{Swatantra~Kumar~Tiwari}
\email{swatantra@allduniv.ac.in (Corresponding Author)}
\affiliation{Department of Physics, University of Allahabad, Prayagraj- 211002, U.P.}

\begin{abstract}
\noindent
In this work, we investigate the drag and diffusion coefficients of $D^0$ meson propagating through a hadronic thermal bath by employing the Fokker–Planck equation within the framework of Tsallis non-extensive statistics. The non-extensive parameter, $q$ accounts for the deviation from equilibrium and provides a more realistic description of the medium that is not perfectly thermalized. The hadronic bath, consisting of various mesonic and baryonic species, is characterized by different mass cutoffs that control the spectral composition of the medium. Our analysis shows that both the drag, $F$ and momentum diffusion coefficients, $\Gamma$ increases with temperature and also increases with increasing $q$ and mass cutoff. The spatial diffusion coefficient, $D_x$ exhibits a decreasing trend with temperature $T$, $q$ and mass cutoff which highlights the significant influence of non-equilibrium effects and hadronic composition on the transport behaviour of $D^0$ meson, offering valuable insights into the thermal and dynamical properties of the hadronic phase in heavy-ion collisions. In this study while calculating the spatial diffusion coefficient, $D_x$ we observe that beyond $q$ = 1.24, $D_x$ goes below to the lower limit proposed in Ads/CFT theory. This suggests that 1.24 is the upper limit of non extensive parameter, $q$.

\end{abstract}


\date{\today}

\maketitle 
\section{Introduction}
\label{intro}

Relativistic heavy-ion collision experiments at major facilities such as RHIC and the LHC offer an exceptional opportunity to explore the properties of strongly interacting matter at extreme temperatures and/or densities. In these high-energy environments, ordinary nuclear matter undergoes a remarkable transformation from a confined hadronic phase to a deconfined state composed of quarks and gluons, known as the quark–gluon plasma (QGP) \cite{Niida:2021wut, Iancu:2012xa, Steinberg:2011dj, Wyslouch:2011zz}. This transition marks a fundamental change in the behaviour of matter governed by Quantum Chromodynamics (QCD). Many studies have been done for extraction of the properties of this deconfined state of matter.
Following the collision, the initially formed hot and dense QGP rapidly expands and cools, eventually hadronizing into a system dominated by interacting hadrons. During this hadronic phase, the medium remains strongly coupled, with significant interactions among its constituents before the system reaches thermal and chemical freeze-out \cite{He:2011qa, Cao:2015hia, Becattini:2014hla}. Analysis of the QGP signatures imprinted in the hadronic phase is of crucial importance.
A number of probes that preserve information from the initial stages of the system’s expansion are: jet quenching \cite{PHENIX:2001hpc} (primary source of high $p_T$ hadrons), direct photons \cite{PHENIX:2008uif} and heavy quarks. Because their masses are much larger ($m_c$=1.3 GeV and $m_b$= 4.3 GeV) than those of light quarks and also greater than the temperature of the system, therefore their thermal relaxation during the subsequent evolution of the fireball occurs more slowly than that of light quarks. The fact can be exploited and thus makes them excellent probes. For these reasons, there is much more interest in heavy mesons formed after the hadronization of charm and bottom quarks.

Electrons originating from the semileptonic decays of heavy mesons act as one of the probes of heavy mesons drag and diffusion. Experimental investigations at RHIC \cite{PHENIX:2006iih} and the LHC \cite{Masciocchi:2011fu} have examined the nuclear suppression factor and elliptic flow of these electrons at nearly vanishing baryochemical potential. The nuclear suppression factor and also the elliptic flow for D mesons has been extracted by the ALICE Collaboration \cite{ALICE:2012ab, ConesadelValle:2012pfy}. These physical observables are closely related to the transport properties of heavy hadrons, which are strongly influenced by their interactions with the surrounding medium. The present work addresses the interaction of a $D^0$ meson with a thermal hadronic medium in a temperature domain relevant for heavy ion phenomenology.
The abundance of heavy mesons remains small within the considered temperature range and do not significantly influence the bulk properties of the medium. Thus, their thermal production is negligible in below critical temperature range. Consequently, the drag and diffusion coefficient of heavy mesons with the thermal hadrons can be evaluated by using elastic interactions \cite{Ghosh:2011bw}.

The evolution of heavy mesons in a hadronic medium is governed by the Boltzmann Transport Equation (BTE). Under the assumption of small momentum transfer in collisions between the heavy meson and the constituents of the medium, the BTE can be reduced to the Fokker Planck Equation (FPE). In this framework, the interaction of heavy mesons drives their evolution with the surrounding medium particles, characterized by transport coefficients such as drag and diffusion which act as the inputs to the Fokker Planck equation \cite{Bhattacharyya:2017sqf}.
The drag coefficient quantifies the average rate of momentum loss due to dissipative forces arising from successive scatterings with the medium constituents. While the momentum diffusion coefficient describes the stochastic broadening of the particle’s momentum distribution, reflecting the random fluctuations induced by microscopic collisions. On the other hand, spatial diffusion coefficient  governs the rate at which the particle spreads through space and is inversely related to the strength of its coupling with the medium.

Recent studies have utilized a range of QCD-inspired theoretical approaches to investigate the temperature dependence of drag and diffusion coefficients for hadrons and heavy mesons. In reference \cite{Laine:2011is}, Heavy Meson Chiral Perturbation Theory was employed to analyze diffusion coefficient, while reference \cite{He:2011yi} examined transport coefficients using empirical elastic scattering amplitudes of $D$ mesons with thermal hadrons. Additionally, effective field theory–based calculations of drag and diffusion were carried out in references \cite{Ghosh:2011bw, vanHees:2004gq}. In reference \cite{He:2011yi}  vacuum scattering amplitudes derived from effective Lagrangians are utilized to study the drag and diffusion coefficients. More recently, transport properties have also been explored within the Color String Percolation Model (CSPM) \cite{Goswami:2022szb} and the Van der Waals Hadron Resonance Gas (VDWHRG) framework \cite{Goswami:2023hdl}. Traditionally, these studies are carried out assuming a medium in perfect thermal equilibrium, described by Boltzmann–Gibbs statistics. However, experimental observations indicate that the particle momentum spectra in heavy-ion collisions often exhibit power-law tails, suggesting significant non-equilibrium effects. To capture such deviations, the Tsallis non-extensive statistics has emerged as a powerful generalization of conventional thermodynamics \cite{Akhil:2023xpb, Singh:2023agf}. The non-extensive parameter $q$ quantifies the degree of deviation from equilibrium, with $q=1$ recovering the Boltzmann–Gibbs limit and $q>1$ representing systems with long-range correlations and memory effects \cite{Tsallis:1987eu, Kyan:2022eqp, Zheng:2015tua, Singh:2023agf}.
In this framework, the non-extensive parameter $q$ modifies the particle distribution functions, directly affecting the drag and diffusion coefficients. Hence, the study under Tsallis statistics reveals key insights into the non-equilibrium transport behaviour of hadrons in a thermal hadronic medium.

The manuscript is organized as follows: In Section~\ref{formulation} we outline the process of determining the drag and diffusion coefficients using the Fokker-Planck equation within the Tsallis distribution function. In Section~\ref{RD}, we have done a comprehensive discussion of the results obtained. Lastly, in Section~\ref{summary}, we provide a succinct summary of the study.

\section{Formulation}
\label{formulation}
\noindent

\label{gamma}

The heavy flavour hadrons are significantly heavier than the constituents of the thermal bath mainly consists of pions. Consequently, due to their much greater mass than the medium constituents and the temperature of the thermal bath, we can treat it as a heavy brownian particle in a thermal bath of light hadronic medium. Thus the transfer of momentum by random kicks is very small compared to that of initial momentum of the particle. The Boltzmann transport equation in this case reduces to a much simpler Fokker-Planck equation in which we do not need to solve the kinetic equation for the particle distribution function $f_c$ and more convenient for the heavy system \cite{Torres-Rincon:2012sda}. The transport coefficients can be obtained by moments of collision integrals. The Fokker-Planck equation in three dimension is given by \cite{Abreu:2011ic},

\begin{eqnarray}
\label{eq1}
 \frac{\partial f_c}{\partial t}= {F}\vec{\nabla_p} . (\vec{P}
f_c) + \Gamma\nabla^2_p f_c
\end{eqnarray}

Here $F$, $\Gamma_0$, $\Gamma_1$ are scalar coefficients and do not depend on 
$\vec{P}$ which means $F(P^2) = F$  and  $\Gamma_0(P^2)$= $\Gamma_1(P^2)$. The first term on the right-hand side corresponds to the drag force, representing the systematic loss of momentum due to interactions with the surrounding. The second term involves the momentum diffusion, characterizing stochastic momentum broadening caused by random scattering with medium constituents.

The standard diffusion equation is given as,

\begin{equation}
 \label{eq2}
  \frac{\partial C}{\partial t}= -{\mu}\nabla . (C \vec{F}) + D\nabla^2 C
\end{equation}

where $\mu$= mobility, {C} = concentration of a solute, {D}= diffusion coefficient, $\vec{F}$= external force.

On comparing the above vector equations we get {F} as friction coefficient or drag and $\Gamma$ as diffusion coefficient which characterize the dissipative and fluctuating interactions with the medium. 

The {F} and ${\Gamma}$ are the moment of transition rates in Fokker-Planck equation \ref{eq1} and is given by~\cite{Ozvenchuk:2014rpa}-

\begin{equation}
 \label{eq3}
     F_i(\vec P)= \int K_i w(\vec P, \vec K)\, d\vec K
\end{equation}

The momentum-space diffusion matrix which accounts for broadening of momentum distribution is:

\begin{equation}
 \label{eq4}
  \Gamma_{ij}(\vec{P})=\frac{1}{2} \int K_i K_j w(\vec{P}, \vec{K})\, d\vec K,
\end{equation}
where ${w(\vec{P}, \vec{K})}$ is transition rate or collision rate, $\vec P$ is initial momenta, $\vec K$  is transferred momenta, $\vec P - \vec K$ is final momenta.\cite{Tolos:2016slr}

In the ideal scenario of a homogeneous and isotropic thermal bath, the coefficients $F_i$ and $\Gamma_{ij}$ which depends solely on the particle momentum $\vec{P}(P_i)$, can be conveniently expressed in terms of three independent scalar functions $F$, $\Gamma_0$ and $\Gamma_1$ as 
$ F_i(\vec P)=F(P^2)P_i $ and $\Gamma_{ij}(\vec P)=\Gamma_0(P^2)\Delta_{ij}+\Gamma_1(P^2)\frac{P_i P_j}{P^2}$
where $\Delta_{ij} = \delta_{ij}-\frac{P_i P_j}{P^2}$ and $\Delta_{ij}\Delta^{ij}=2$ \cite{Abreu:2011ic}.

Now Fokker Planck in one dimension is given by,

\begin{equation}
 \label{eq5}
  \frac{\partial f_c}{\partial t}= {F}\frac{\partial(pf_c)}{\partial p} + \Gamma \frac{\partial^2 f_c}{\partial p^2}
\end{equation}

The above equation also known as Rayleigh's equation which describes momentum distributon equation for a brownian particle.

Now using initial conditions ${f_c}(p, t=0) = \delta (p-p_0)$
The analytic solution of equation \ref{eq5} is given by~\cite{Abreu:2011ic},

\begin{equation}
 \label{eq6}
  f_c(p,t) = \left[\frac{F}{2\pi \Gamma} (1- e^{-2Ft})\right]^{\frac{-1}{2}} \exp\left[{\frac{-F (p-p_0 e^{-Ft})^2}{2\Gamma (1-e^{-2Ft})}}\right]
\end{equation}

 The term $<p>=p_0 e^{-Ft}$ represents the exponential decay of the mean momentum of the particle due to the action of the drag force, which continuously dissipates its initial momentum through successive interactions with the surrounding medium. Simultaneously, the factor $<p^2>-<p>^2=
 \frac{\Gamma}{F}(1-e^{-2Ft})$  in the denominator governs the broadening of the momentum distribution. 
 
 Here {F} and ${\Gamma}$ are not independent but related by Einstein's fluctuation-dissipation relation

\begin{equation}
\label{eq7}
  \Gamma = FmT  
\end{equation}

Within the Fokker-Planck approach, spatial diffusion coefficient $(D_x)$ can be obtained from the drag coefficient with $m$ is the mass of heavy particle
 
 \begin{equation}
 \label{eq8}
     D_x= \frac{\Gamma}{{m^2}{F^2}} = \frac{T^2}{\Gamma} = \frac{T}{{m}{F}}
 \end{equation}

\begin{figure}[h]
\centering

\begin{minipage}{0.49\textwidth}
    \centering
    \includegraphics[width=\linewidth]{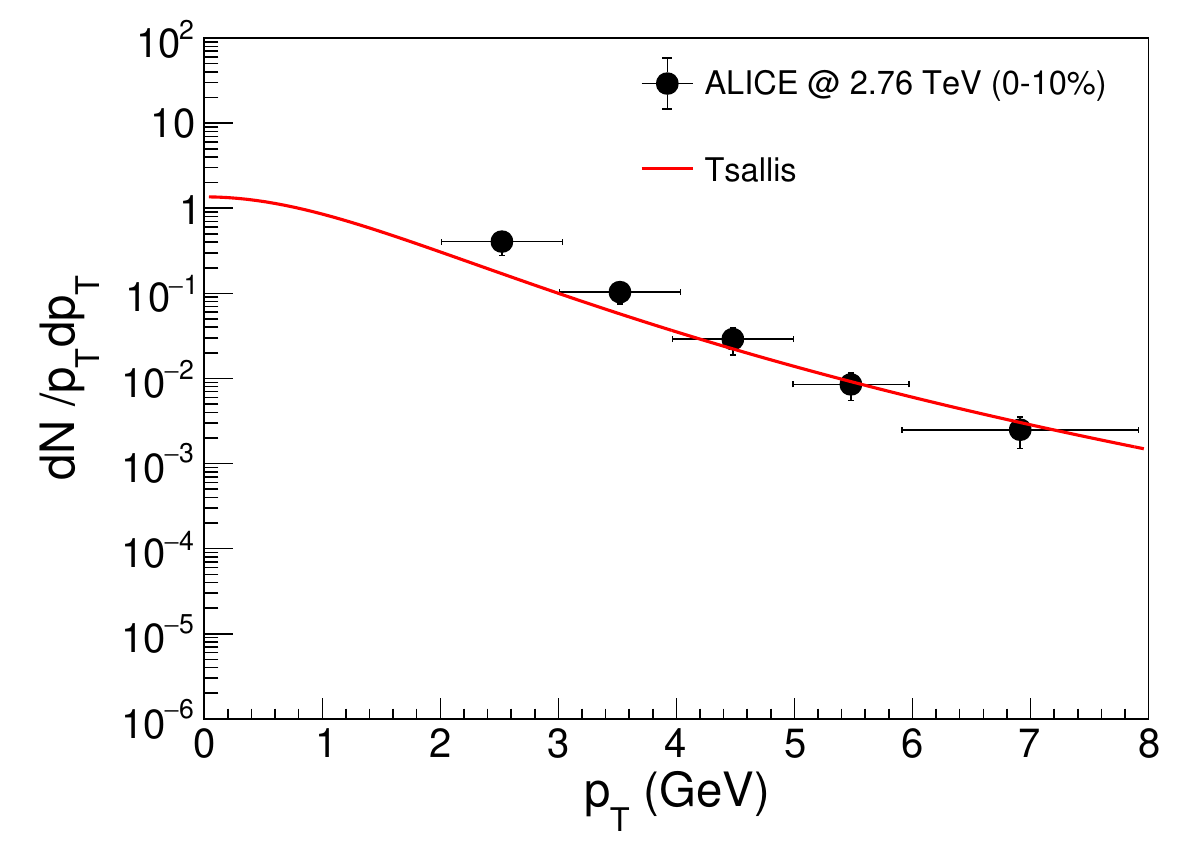}
    \caption{Transverse momentum spectra of $D^0$ meson at $q$=1.16 and $T$=180 MeV in Tsallis statistics and the experimental data shown is taken from ALICE @2.76 TeV for Pb-Pb collisions at (0-10\%) centrality.}
    \label{figa}
\end{minipage}
\end{figure}

Now if $p(t)$ is survival probability = $e^{-wt}$ where $w$ is transition rate or collision rate that depends on velocity $v$. Then 
$p(t)dt$ = survival probability for $dt$ time interval = $we^{-wt}dt$. 
The relaxation time or mean time is $\tau$ = $\bar{t}$ = $\int_{0}^{\infty} t.p(t)dt $ = $\int_{0}^{\infty} wt.e^{-wt}dt $ = $\frac{1}{w}$
OR 
$\tau^{-1}=w$. For the process $a(p_a)+b(p_b) \rightarrow  a(p_c)+b(p_d)$, if $n_a$ is number of particles per unit volume with relative velocity $ v_{ab}$ contained in volume $v_{ab}dtdA$ , then relative incident flux = $n_a v_{ab}$. Number of scattered particles will give collision probability as \cite{reif1965fundamentals}-

\begin{equation}
\label{eq9}
    w=\tau^{-1}=\frac{(n_av_{ab})\sigma_{ab}(n_bd^3r)}{n_ad^3r}
\end{equation}

\begin{equation}
    \label{eq10}
\tau^{-1}=n_bv_{ab}\sigma_{ab}
\end{equation}

Average relaxation time of $a^{th}$ particle
\begin{eqnarray}
\label{eq11}
\tilde{\tau_a}^{-1}= \frac {\int{\frac{d^3p_a}{(2\pi)^3}\tau^{-1}f(E_a)}} {\int{\frac{d^3p_a}{(2\pi)^3}f(E_a)}}
\end{eqnarray}

Putting the value of $\tau^{-1}$ from equation \ref{eq10} we get,
\begin{equation}
\label{eq12}
    \tilde{\tau_a}^{-1}= \sum_{b} \frac {\int{\frac{d^3p_a}{(2\pi)^3}\frac{d^3p_b}{(2\pi)^3}\sigma_{ab}v_{ab}f(E_a)f(E_b)}} {\int{\frac{d^3p_a}{(2\pi)^3}f(E_a)}}
\end{equation}
which is reduced to the form,
\begin{equation}
    \label{eq13}
\tilde{\tau_a}^{-1}=\sum_{b} n_b<\sigma_{ab}v_{ab}>
\end{equation}
Here $n_b = \int \frac{d^3p_b}{(2\pi)^3}f(E_b)$  is the number density and $f(E_b)$ is the distribution function of $b^{th}$ hadronic species.

General definition of thermal average in non extensive statistics
\begin{equation}
\label{eq14}
    <\sigma_{ab}v_{ab}>=\frac{\int \frac{d^3p_a}{E_a}\frac{d^3p_b}{E_b}e_q^{-E_a/T}e_q^{-E_b/T}\sigma v_{ab}}{\int  \frac{d^3p_a}{E_a}\frac{d^3p_b}{E_b}e_q^{-E_a/T}e_q^{-E_b/T}}
\end{equation}

The elastic scattering cross section $\sigma$ for mesons and baryons are taken as 10 mb and 15 mb respectively at zero baryon density \cite{Ozvenchuk:2014rpa, Goswami:2023hdl}.

Here in this study we are employing the  Tsallis distribution funtion (in the local rest frame) which is written as,  
\begin{equation}
f(E_a)=[1+(q-1)\frac{E_a}{T}]^{-q/(q-1)}=e_q^{-E_a/T}.
\end{equation}

The invariant particle number is given as,
\begin{equation}
d^3N= gV \frac{d^3p}{(2\pi)^3}f(E_a)
\end{equation}

where $g$ is spin-isospin degeneracy factor and the above equation will give the number of particles between momentum $p$ to $p+dp$ and $f(E_a)$ is the distribution function. Now putting the value of $f(E_a)$ and $d^3p= p_Tdp_Td\phi dp_z$ where $\frac{dp_z}{E}=dy$ and $E=m_Tcoshy$ into above equation and integrating over $y$ we get,

\begin{equation}
\label{eq00}
\frac{dN}{p_Tdp_T} = \int{\frac{gVm_Tcoshy}{(2\pi)^2}\left[1+(q-1)\frac{m_Tcoshy}{T}\right]^{-q/(q-1)}dy}
\end{equation}

\begin{figure*}[htb]
\hfill
{ 
\begin{minipage}[b]{0.49\textwidth}
\centering \includegraphics[width=\linewidth]{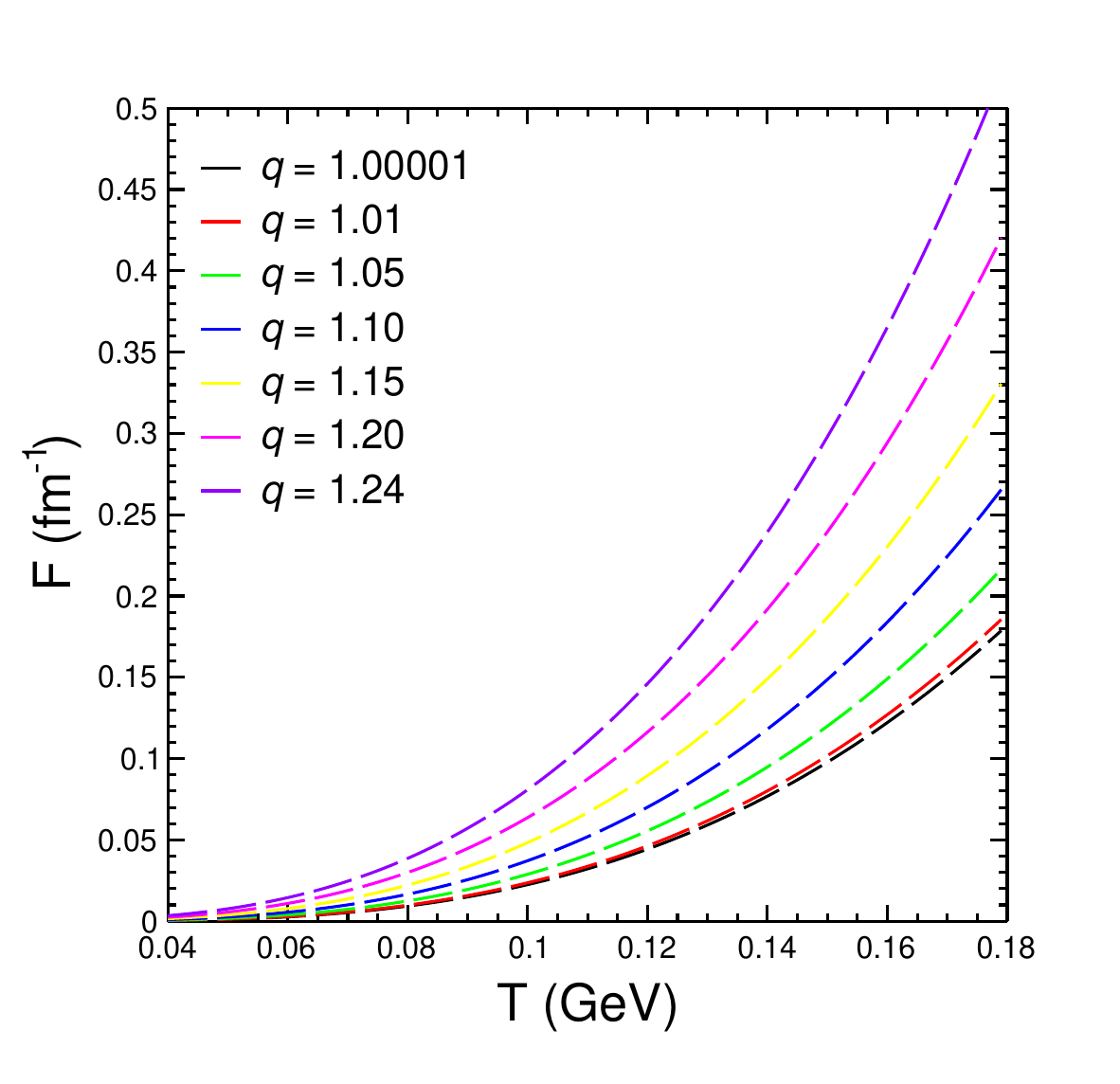}
\end{minipage}}
\hfill
{ 
\begin{minipage}[b]{0.49\textwidth}
\centering \includegraphics[width=\linewidth]{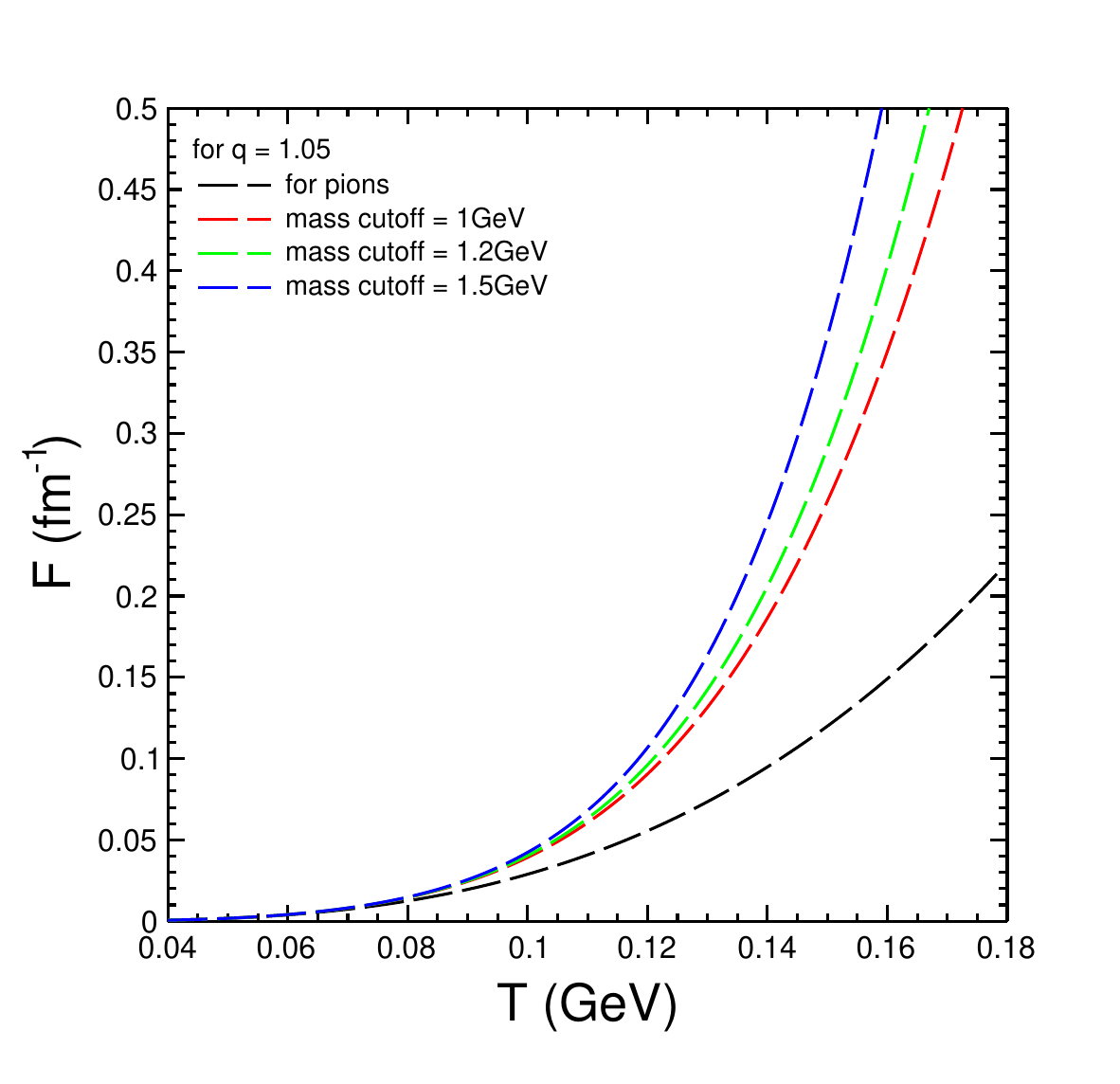}
\end{minipage}}

\caption{Variation of Drag coefficient of $D^0$ meson in pion gas (left panel) for various q values and the drag coefficient of $D^0$ meson in thermal medium contolled by mass cutoffs (right panel) for a fixed $q$ value as a function of temperature.}
\label{fig1} 
\end{figure*}
Now the momentum space volume elements can be written as \cite{Tiwari:2017aon}
\begin{equation}   
\frac{ d^3p_a}{E_a}\frac{d^3p_b}{E_b}=8\pi^2p_ap_bdE_adE_bdcos\theta
\end{equation}

and 
\begin{widetext}
{\large
\begin{equation}
\label{eq15}
v_{ab}=\frac{\sqrt{(E_aE_b-p_a p_b\cos\theta)^2-(m_am_b)^2}}{E_aE_b-p_ap_b \cos\theta}
\end{equation}
}

{\large
\begin{equation}
\begin{aligned}
\label{eq:16}
<\sigma_{ab}v_{ab}>=\frac{\sigma \int{8\pi^2p_ap_bdE_adE_bdcos\theta e_q^{-E_a/T}e_q^{-E_b/T}\times \frac{\sqrt{(E_aE_b-p_ap_bcos\theta)^2-(m_am_b)^2}}{E_aE_b-p_ap_bcos\theta}}}{\int{8\pi^2p_ap_bdE_adE_bdcos\theta e_q^{-E_a/T}e_q^{-E_b/T}}}
\end{aligned}
\end{equation}
}

The relaxation time is related to drag coefficient as $\tau^{-1} = F$ and is given by \cite{reif1965fundamentals}

Thus drag
\begin{equation}
\label{eq17}
  F=\tilde{\tau_a}^{-1}=\sum_{b} n_b<\sigma_{ab}v_{ab}>
\end{equation}

\begin{figure*}[htb]
\hfill
{ 
\begin{minipage}[b]{0.49\textwidth}
\centering \includegraphics[width=\linewidth]{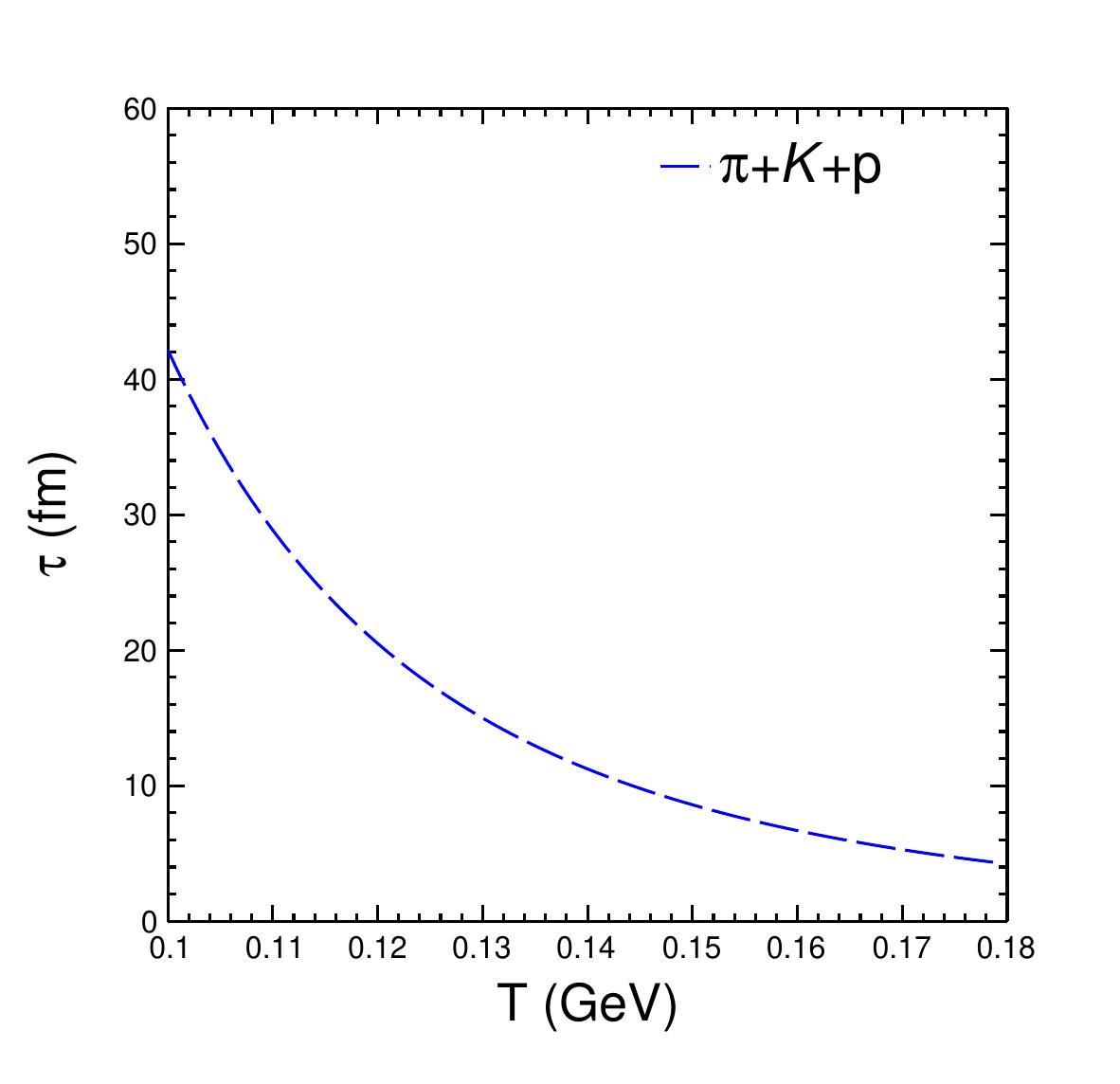}
\end{minipage}}
\hfill
{ 
\begin{minipage}[b]{0.49\textwidth}
\centering \includegraphics[width=\linewidth]{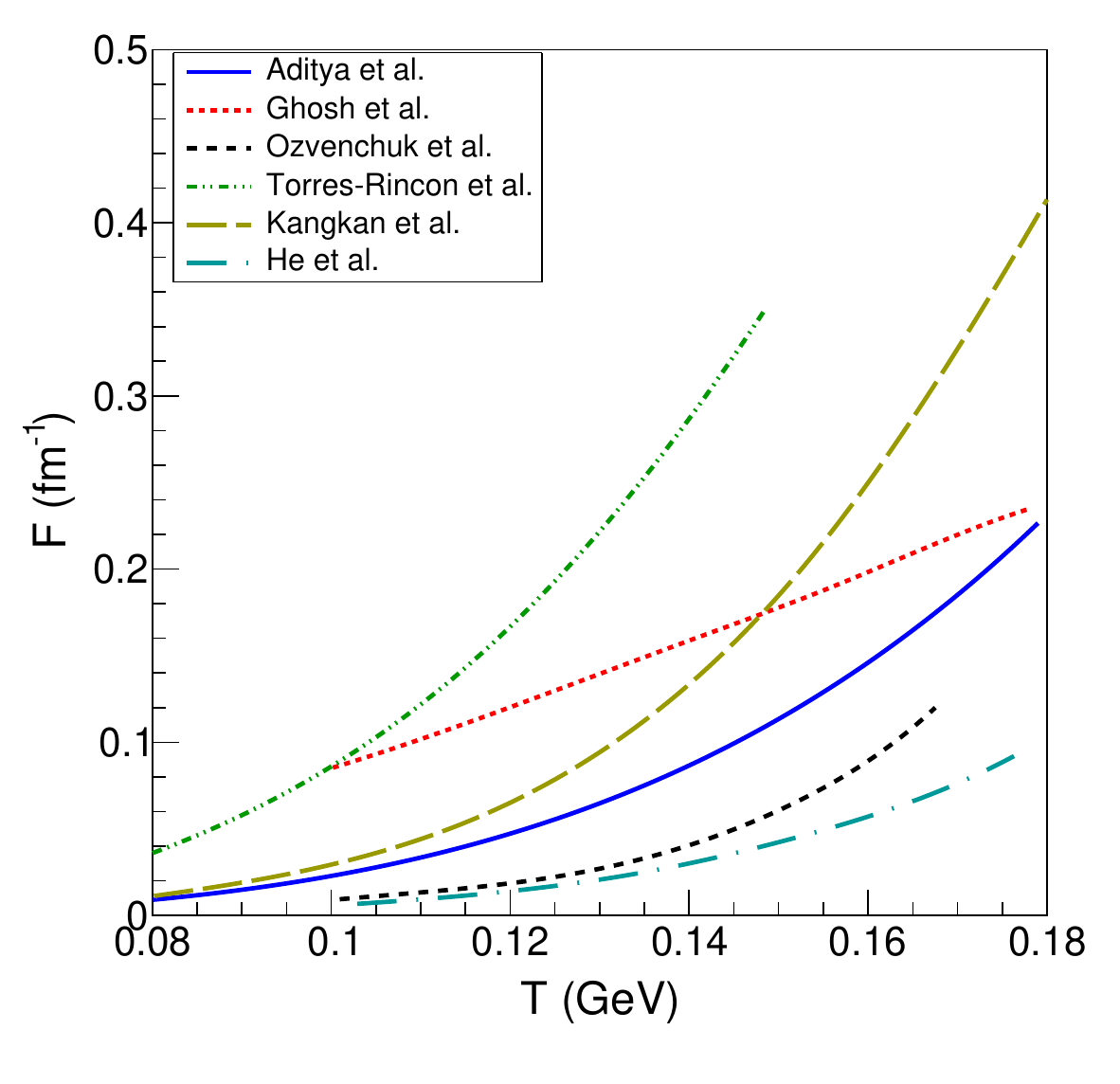}
\end{minipage}}

\caption{Variation of relaxation time of $D^0$ meson with temperature in pion, kaon and proton medium for a fix $q$ value (left panel). The drag coefficients of $D^0$ meson in pion, kaon and proton medium for $q$ = 1.00001 is shown in comparison to different phenomenological models estimates by Torres-Rincon et al. \cite{Torres-Rincon:2021yga} (green dashed dot dot line), Ghosh et al. \cite{Ghosh:2011bw} (red dashed line), Ozvenchuk et al. \cite{Ozvenchuk:2014rpa} (black dashed line), He et al. \cite{He:2011yi} (cyan dashed dot line) and Kangkan et al. \cite{Goswami:2023hdl} (yellow dashed line)
as a function of temperature (right panel).}
\label{fig2} 
\end{figure*}

\begin{figure*}[]
{
\begin{minipage}[b]{0.48\textwidth}
\centering \includegraphics[width=\linewidth]{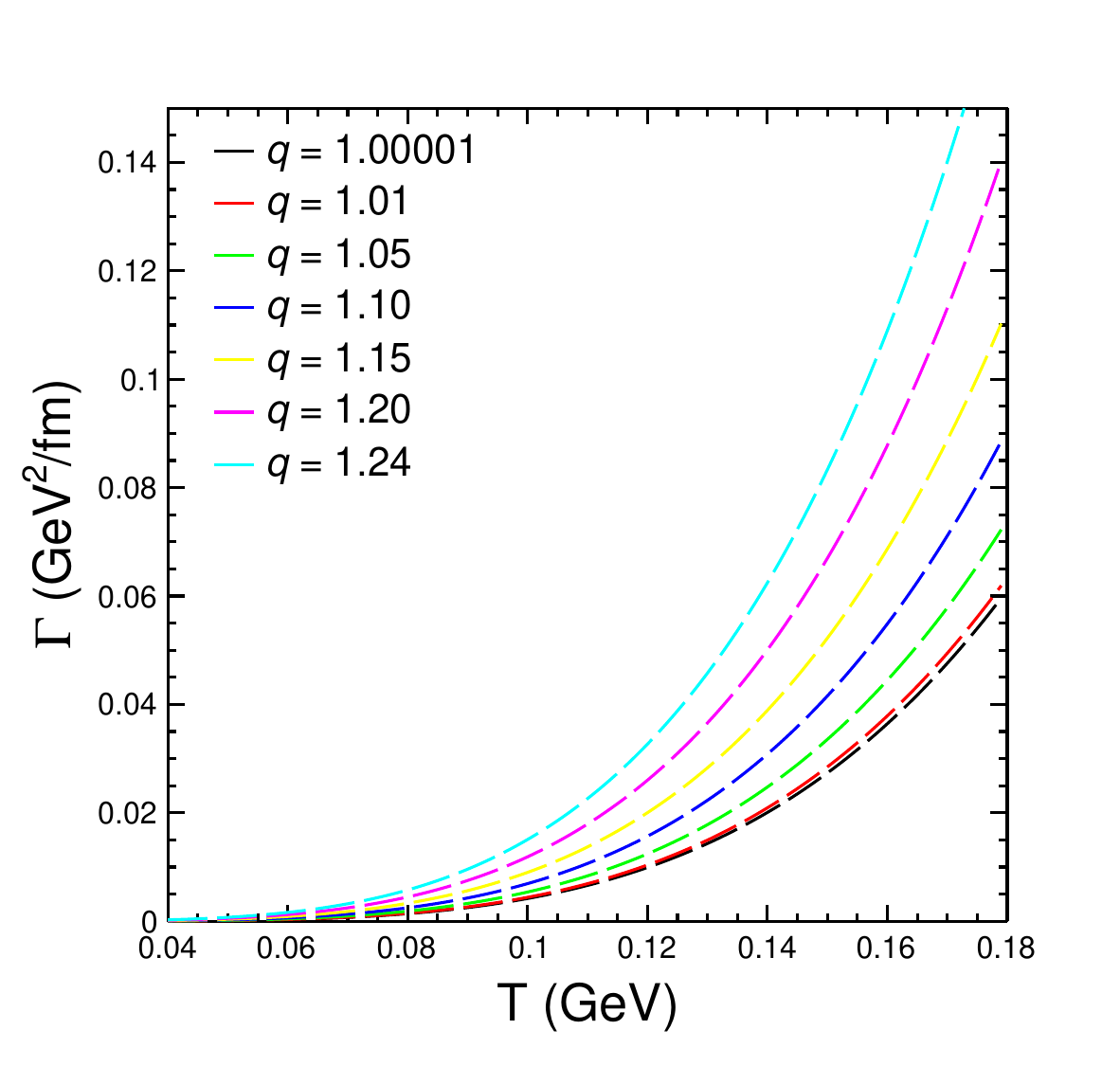}
\end{minipage}}
\hfill
{ 
\begin{minipage}[b]{0.48\textwidth}
\centering \includegraphics[width=\linewidth]{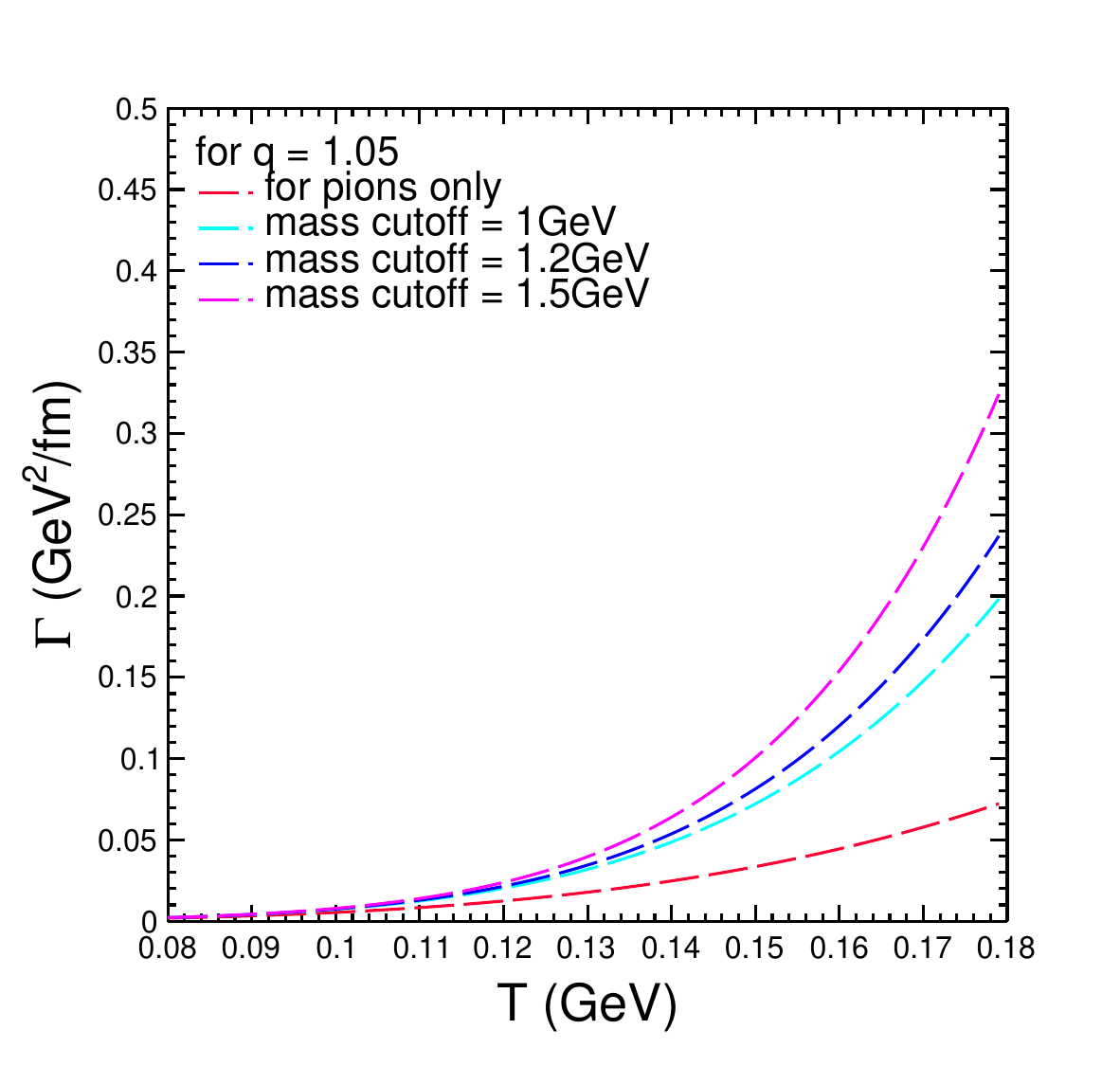}
\end{minipage}}
\caption{Variation of Momentum Diffusion coefficient of $D^0$ meson in pion gas with temperature for various $q$ values (left panel) and different mass cutoffs of medium constituents (right panel).}
\label{fig3} 
\end{figure*}

{\large
\begin{equation}
\begin{aligned}
\label{eq:18}
  F= \sum_{b} n_b\frac{\sigma \int{8\pi^2p_ap_bdE_adE_bdcos\theta e_q^{-E_a/T}e_q^{-E_b/T}\times\frac{\sqrt{(E_aE_b-p_ap_bcos\theta)^2-(m_am_b)^2}}{E_aE_b-p_ap_bcos\theta}}}{\int{8\pi^2p_ap_bdE_adE_bdcos\theta e_q^{-E_a/T}e_q^{-E_b/T}}}
\end{aligned}
\end{equation}
}

Here ${F}$ and ${\Gamma}$ are not independent but related by Einstein's fluctuation-dissipation relation

\begin{equation}
\label{eq19}
 \Gamma=FmT=mT \sum_{b} n_b<\sigma_{ab}v_{ab}>
\end{equation}

{\large
\begin{equation}
\begin{aligned}
\label{eq20}
  \Gamma=mT\sum_{b} n_b\frac{\sigma \int{8\pi^2p_ap_bdE_adE_bdcos\theta e_q^{-E_a/T}e_q^{-E_b/T}\times\frac{\sqrt{(E_aE_b-p_ap_bcos\theta)^2-(m_am_b)^2}}{E_aE_b-p_ap_bcos\theta}}}{\int{8\pi^2p_ap_bdE_adE_bdcos\theta e_q^{-E_a/T}e_q^{-E_b/T}}}
\end{aligned}
\end{equation}
}

\begin{figure}[h]
\centering

\begin{minipage}{0.49\textwidth}
    \centering
    \includegraphics[width=\linewidth]{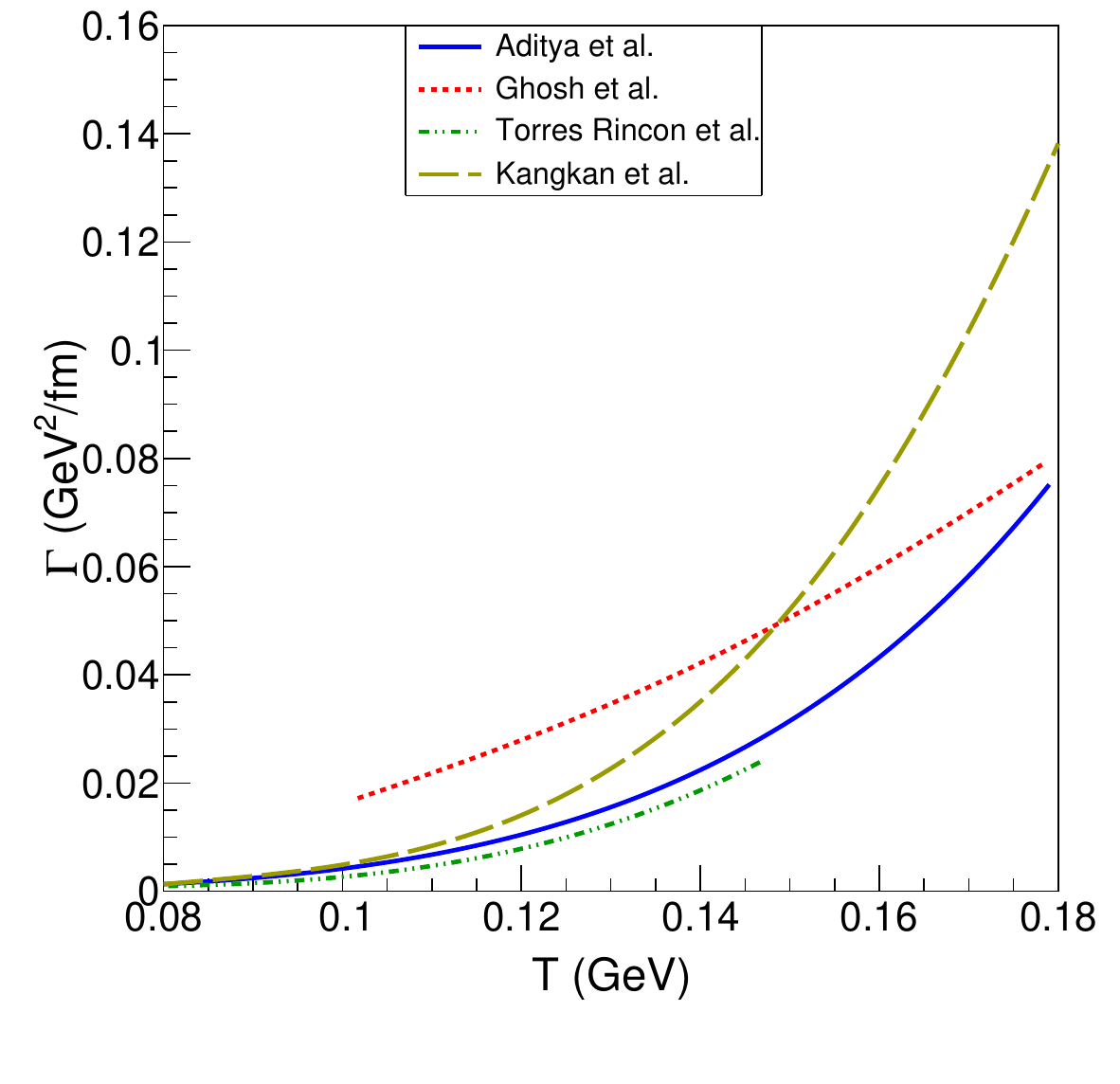}
    \caption{Momentum Diffusion coefficient of $D^0$ meson in pion, kaon and proton medium for $q$ = 1.00001 is shown in comparison to phenomenological models taken from Ghosh et al. \cite{Ghosh:2011bw}(red dotted line), Torres-Rincon et al. \cite{Torres-Rincon:2021yga}, (green dashed dotted dotted line), Kangkan et al. \cite{Goswami:2023hdl} (yellow dashed line).}
    \label{fig4}
\end{minipage}
\end{figure}

\begin{figure*}[]
{
\begin{minipage}[b]{0.48\textwidth}
\centering \includegraphics[width=\linewidth]{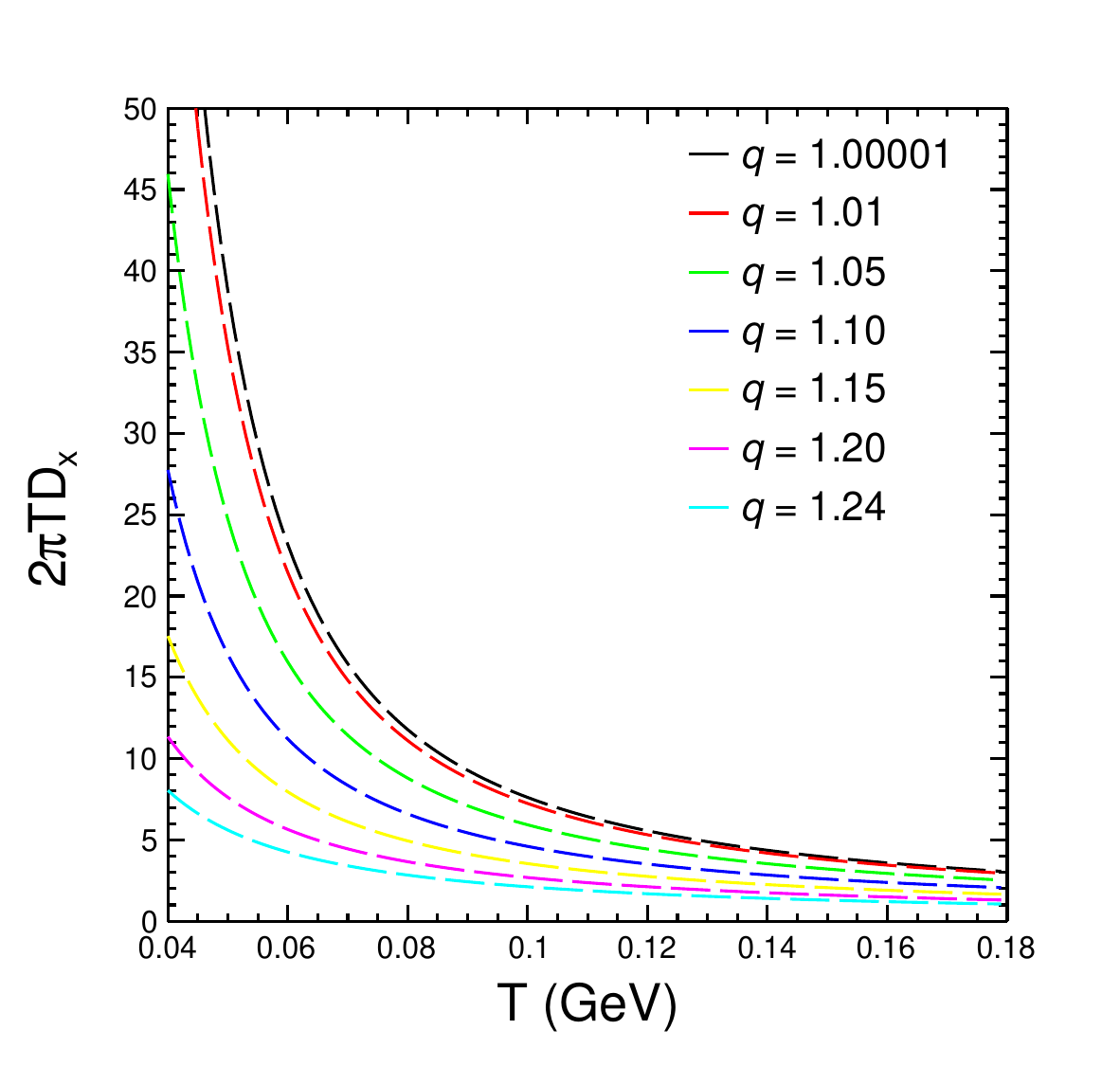}
\end{minipage}}
\hfill
{ 
\begin{minipage}[b]{0.48\textwidth}
\centering \includegraphics[width=\linewidth]{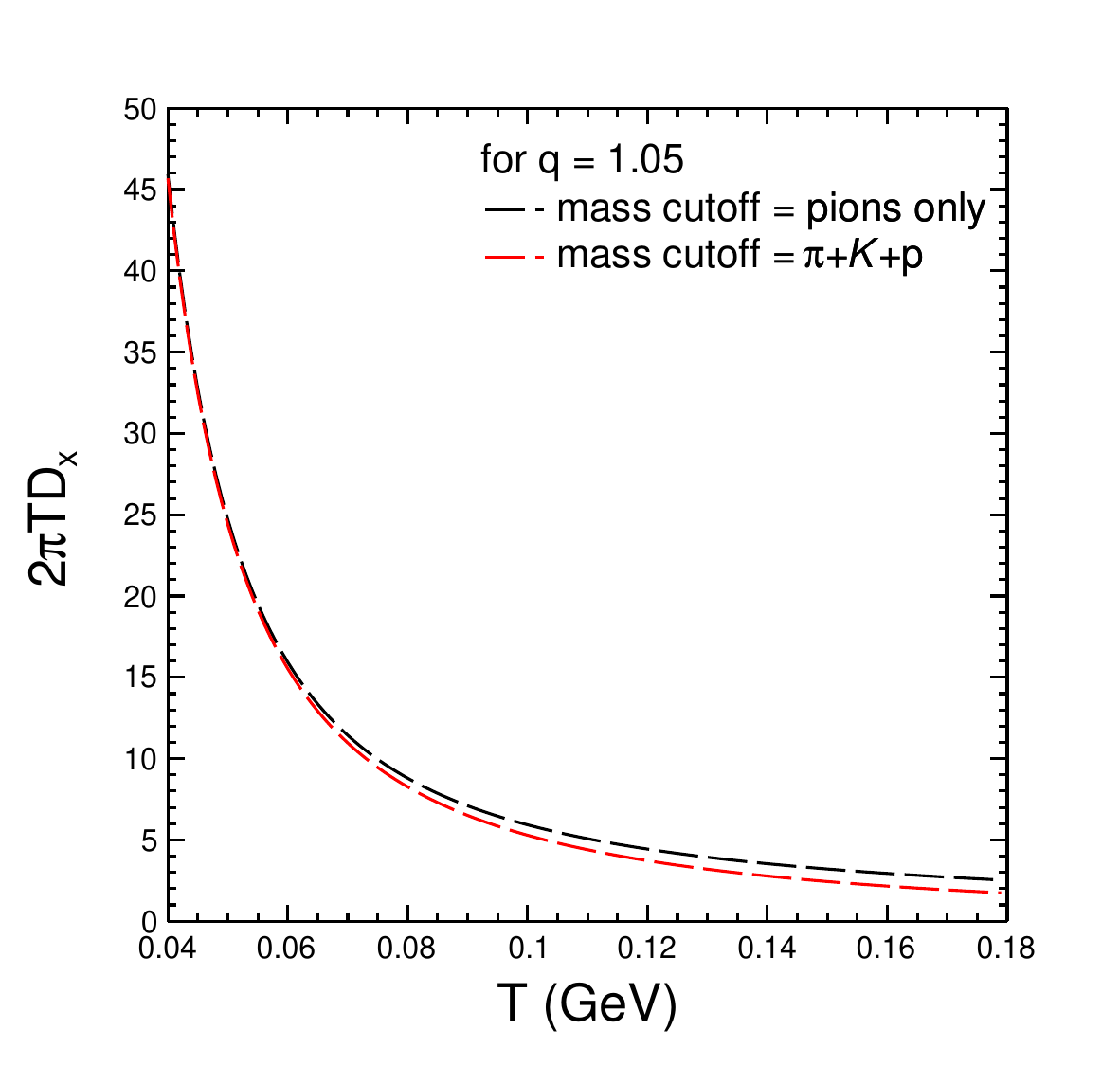}
\end{minipage}}
\caption{Variation of Spatial Diffusion coefficient of $D^0$ meson with temperature for various $q$ values (left panel) and constituent mass of medium is $\pi + K + p$ (right panel).}
\label{fig5} 
\end{figure*}

Within the Fokker-Planck approach, spatial diffusion coefficient ($D_x$) can be obtained from the drag coefficient with $m$ is the mass of heavy particle

\begin{equation}
\label{eq21}
D_x=\frac{T}{mF}=\frac{T}{m \sum_{b} n_b<\sigma_{ab}v_{ab}>}
\end{equation}

{\large
\begin{equation}
\begin{aligned}
\label{eq22}
  D_x=\frac{T} {\displaystyle{m \sum_{b} n_b\frac{\sigma \int{8\pi^2p_ap_bdE_adE_bdcos\theta e_q^{-E_a/T}e_q^{-E_b/T}\times\frac{\sqrt{(E_aE_b-p_ap_bcos\theta)^2-(m_am_b)^2}}{E_aE_b-p_ap_bcos\theta}}}{\int{8\pi^2p_ap_bdE_adE_bdcos\theta e_q^{-E_a/T}e_q^{-E_b/T}}}}}
\end{aligned}
\end{equation}

}
\end{widetext}

\section{Results and Discussions}
\label{RD}

In this section, we present a detailed analysis of the drag and diffusion coefficients of $D^0$ meson propagating through a hadronic thermal bath as functions of temperature across the medium along with varying mass of medium composition. 
The transport coefficients have been evaluated using the Fokker Planck formalism within the framework of non extensive Tsallis statistics, which provides a generalized description of systems that deviate from ideal equilibrium conditions. The non-extensive parameter $q$ quantifies the degree of deviation from equilibrium and thus related to relaxation time of the system. Therefore, $q$ parameter can serve as a crucial probe for the medium response to it.
The interactions of $D^0$ with the surrounding medium or the thermal bath are influenced by the inclusion of different heavy mass particles into the medium. To account for these contributions, different hadronic mass cutoffs have been implemented which provide us to check the sensitivity of drag and diffusion on the constituent medium and interaction with them.

In figure~\ref{figa} we have shown the transverse momentum spectra of $D^0$ meson calculated using Tsallis statistics given by Eq. \ref{eq00} at LHC energy of 2.76 TeV for Pb-Pb collisions \cite{ALICE:2012ab, Fontoura:2025fsv}. We have compared the calculated results with the experimental data where temperature $T$ is taken as 180 MeV and $q$ as 1.16 which describes the experimental data. We can see that with the considered parameter Tsallis statistics describes the transverse momentum spectra well particularly in high $p_T$ range.

\begin{figure}[h]
\centering
\begin{minipage}{0.49\textwidth}
\centering
    \includegraphics[width=\linewidth]{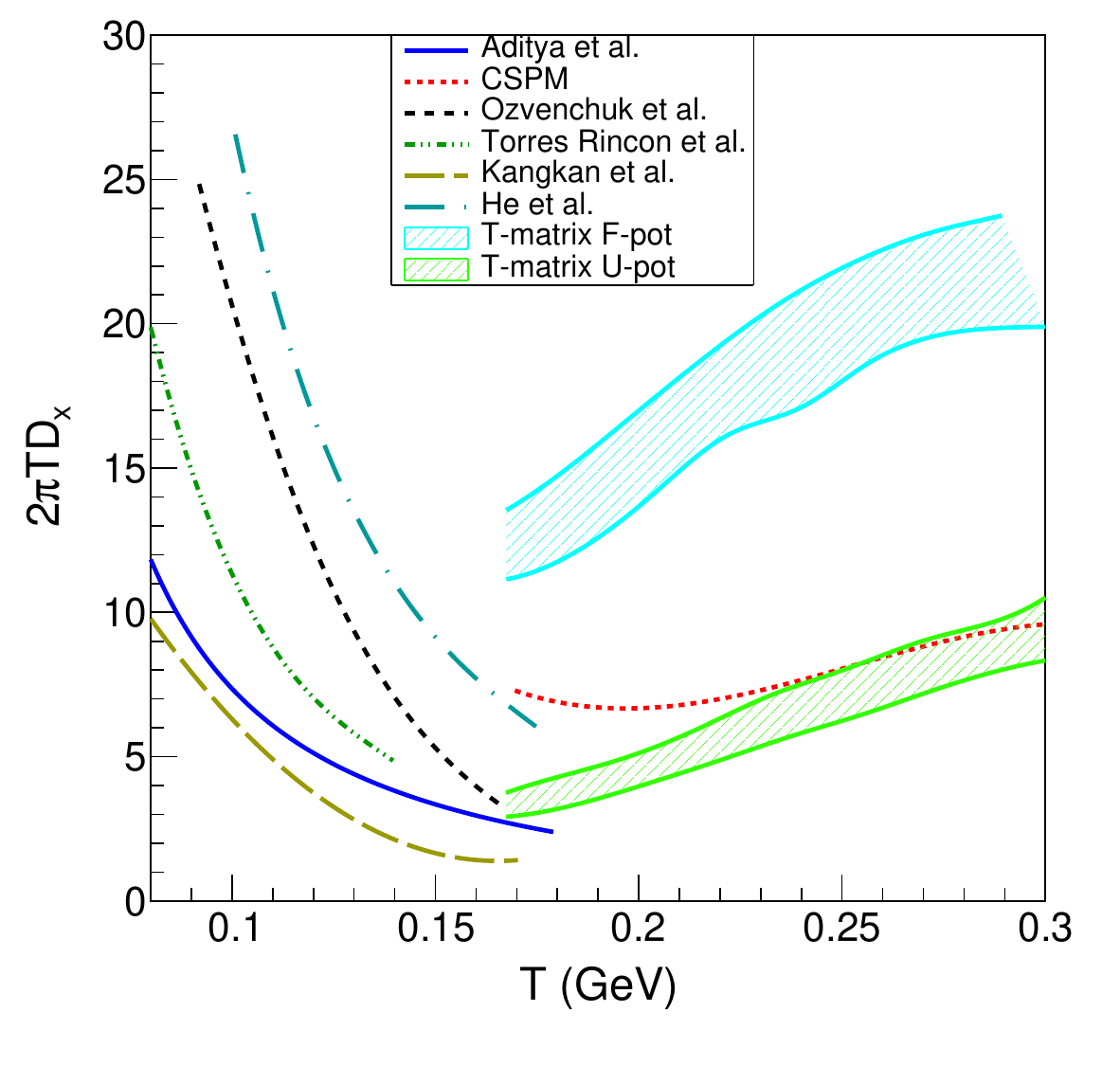}
    \caption{Spatial Diffusion coefficient of $D^0$ meson in $\pi + K + p$ medium for $q$ = 1.00001 is shown and compared with various models such as  Ozvenchuk at al. \cite{Ozvenchuk:2014rpa} (black dotted line), Torres-Rincon et al. \cite{Torres-Rincon:2021yga} (green dashed dotted dotted line), Kangkan et al. \cite{Goswami:2023hdl} (yellow dashed line), He et al. \cite{He:2011yi} (cyan dashed dotted line) in the hadronic phase while \cite{Riek:2010fk} (cyan and spring band from T-matrix calculation using $U$- and $F$-potentials), color string percolation model (CSPM) \cite{Goswami:2022szb} (red dotted line) in QGP matter.}
    \label{fig6}
\end{minipage}
\end{figure}

In Figure~\ref{fig1}, we have shown the temperature dependence of the drag coefficient for various values of the Tsallis non-extensivity parameter $q$ by using Eq.(\ref{eq:18}). We notice the drag coefficient increases exponentially with temperature and a systematic increase in drag while increase in $q$ values and mass cutoffs which constitute the thermal bath. 
The rise in drag coefficient with increasing temperature reflects the growing strength of interactions between the propagating particle and the constituents of the hadronic medium. As the temperature increases, the thermal population and average momentum of hadrons become significantly higher, leading to an enhanced rate of collisions and stronger momentum dissipation.

The $q$ parameter shows the non-equilibrium effects which enhance the probability of energetic scatterings between the particle and medium constituents, resulting in higher momentum transfer and consequently larger drag. The increase of the drag ($F$) with $q$ therefore shows a good influence of non equilibrium dynamics on the transport behaviour of the hadronic system. We also estimated the $D^0$ meson drag coefficient, in a thermal bath with different mass cutoffs by using the eq.(\ref{eq:18}). For only pion gas the magnitude of $F$ is very low while in case of inclusion of higher masses into the system the drag increases sharply with temperature. Which means by taking the more higher mass particles into the medium constituent, the drag will become more sensitive to temperature.

In Fig.~\ref{fig2} (left) the values of relaxation time, $\tau$ of $D^0$ meson in $\pi + K + p$ medium as a function of temperature for $q$=1.00001 is shown.
At T=180 MeV we obtain the value of $\tau$ as 4-5 fm. Additionally, we observe the relaxation time increases with decreasing temperature which is the inverse behaviour of the drag coefficient.

In the right panel of fig.~\ref{fig2} we have shown a comparison of drag coefficient obtained from our calculation in $\pi+K+p$ medium with different phenomenological models. In Ref.\cite{Ghosh:2011bw} Ghosh et al. evaluated the drag of open charm mesons in a hot hadronic medium consisting of pions, nucleons, kaons and eta using  the framework of effective field theory. Torres-Rincon et al. \cite{Torres-Rincon:2021yga} derive the off-shell Fokker-Planck equation using Kadanoff Baym approach calculated the heavy-flavor transport coefficients. Ozvenchuk et al.\cite{Ozvenchuk:2014rpa} also computed the drag of heavy meson. He at al.\cite{He:2011yi} calculated the drag of open charm meson in hadronic matter using empirical elastic scattering amplitudes. Kangkan et al.\cite{Goswami:2023hdl} computed the transport coefficients of open charmed meson in a hadronic medium using van der Waals hadron resonance gas model. We also find that the temperature dependence of the drag coefficient in other models shows a similar behavior to our results, which supports the reliability of our calculations.

Figure~\ref{fig3} illustrates the variation of the momentum diffusion coefficient of the $D^0$ meson with temperature for different values of the Tsallis $q$ parameter and the hadronic medium mass using Eq.(\ref{eq20}). We see an increase in momentum diffusion along with temperature, $q$ and mass cutoffs. On increasing temperature the frequency and strength of particle collisions increases resulting in higher momentum broadening.
The trend we observe for momentum diffusion coefficient are in line with the other model estimations by Ghosh et al.\cite{Ghosh:2011bw}, Torres-Rincon et al.\cite{Torres-Rincon:2021yga} and Kangkan et al.\cite{Goswami:2023hdl} as shown in fig.~\ref{fig4}.

In Fig.~\ref{fig5} we have shown the variation of spatial diffusion coefficient of $D^0$ meson against temperature for various $q$ values and various constituent mass of the medium using Eq.(\ref{eq22}). Here we have computed the spatial diffusion coefficient normalized by the de Broglie wavelength $({2\pi T})^{-1}$.
A decreasing trend is observed with increasing temperature, $q$ and mass cutoff similarly observed in references \cite{He:2011yi,Torres-Rincon:2021yga,Ozvenchuk:2014rpa}. This is due to increase in number density and
interaction strength of the medium as we increase temperature and medium constituent mass which results in lowering the mean free path of the particle, leading to a lower spatial diffusion coefficient. Here we find that the spatial diffusion coefficient decreases with $q$, which goes in line with the fact that as we move towards more non equilibrium conditions of a system, the number of collisions increases and hence the spatial diffusion coefficient decreases.

This behaviour is consistent with the corresponding rise in the drag and momentum diffusion coefficients, since the spatial diffusion coefficient $D_x$ is inversely related to these quantities (typically $D_x$=$\frac{T^2}{\Gamma}$=$\frac{T}{mF}$).

Like viscous coefficients, such as the shear and bulk viscosities, when normalized
to the entropy density, show an extremum close to the phase transition temperature \cite{Dobado:2008vt, Chakraborty:2010fr, Khvorostukhin:2010aj, Dobado:2012zf}. Similarly,  the behaviour of the adimensional quantity $2\pi TD_x$ near the deconfinement temperature  has also an extremum value. The AdS/CFT calculation provide the lower bound value of $2\pi TD_x$ near critical temperature $T_c$ as 1 \cite{Herzog:2006gh, Casalderrey-Solana:2006fio}. Using this lower limit we have extracted the upper limit of non extensive parameter, $q$ which turns out to be 1.24.

In fig.~\ref{fig6}, we have displayed the comparision of the calculations of the spatial diffusion coefficient of $D^0$ meson with the results obtained in various models. Our results for the hadronic sector align well with the results obtain by Torres Rincon et al.\cite{Torres-Rincon:2021yga}, Ozvenchuk et al.\cite{Ozvenchuk:2014rpa}, He et al.\cite{He:2011yi}, Kangkan et al.\cite{Goswami:2023hdl} align well with our result. The presence of a minimum around the critical temperature can be understood as a consequence of the emergence of the deconfined medium. In the partonic phase, the result obtained from the T-matrix calculation using $U$- and $F$-potentials \cite{Riek:2010fk} and Color String Percolation Model (CSPM) \cite{Goswami:2022szb} show an increasing trend with increasing temperature. This is due to the fact that at higher temperatures, the partons will be asymptotically free and results in weakening of strong force causing spatial diffusion to increase with temperature.

\section{Summary and Outlook}
\label{summary}
In summary, the drag and diffusion coefficients of $D^0$ meson have been investigated within the Fokker–Planck framework using Tsallis non-extensive statistics, accounting for the non-equilibrium nature of the hadronic medium produced in relativistic heavy-ion collisions.

\begin{enumerate}

\item The drag coefficient exhibits an exponential rise with temperature, suggesting that the interaction between the particle and the medium becomes increasingly stronger at higher temperatures. This reflects the enhanced coupling strength of the system. Such increase signifies more frequent and intense scattering processes within the medium.

\item The drag coefficient also shows an increasing trend with the non-extensive parameter $q$, indicating the significant role of non-equilibrium dynamics in the system. As $q$ deviates from unity, the medium departs from thermal equilibrium, leading to enhanced momentum transfer between the particles. This rise with $q$ highlights how non-extensivity contribute to greater momentum dissipation and stronger medium interactions.

\item An increase in the hadronic mass cutoff leads to a higher drag coefficient. The presence of these massive states enhances the overall interaction rate and scattering processes. Consequently, the medium becomes denser and more strongly coupled, resulting in greater resistance to particle motion and increased drag.

\item The momentum diffusion coefficient also exhibits a similar rising trend with temperature, the non-extensive parameter $q$ and the hadronic mass cutoff. This trend indicates intensified random momentum exchanges between the particle and the medium. The combined effects of higher temperature, stronger non-equilibrium dynamics, and heavier states contribute to enhanced momentum transfer within the system.

\item In contrast to the drag and momentum diffusion coefficients, the spatial diffusion coefficient decreases with increasing temperature, non-extensive parameter, $q$ and hadronic mass cutoff. This decline indicates that the particles mobility within the medium becomes progressively limited under such conditions. The stronger interactions at higher temperature and collision with heavier masses lead to reduced spatial transport. Consequently, the meson experiences higher degree of localization within the medium.

\end{enumerate}
 
\section*{ACKNOWLEDGEMENTS}
AKS is grateful to University Grants Commission (UGC), New Delhi for providing research fellowship. SKT acknowledges the financial support of the seed money grant provided by the University of Allahabad, Prayagraj. Authors acknowledge Dr. Sabyasachi Ghosh for the fruitful discussions.

\bibliographystyle{unsrt}
\bibliography{drag_diffusion_Notes.bib}

\end{document}